\documentclass[preprint,aps,floatfix,nofootinbib,superscriptaddress]{revtex4-1}
\pdfoutput=1

\usepackage{amsmath, amssymb, amsfonts, amsthm, latexsym, epsfig, mathrsfs, xcolor, bbm, subfig, caption}

\usepackage[inline]{enumitem}

\usepackage{setspace}
\usepackage[marginal, multiple]{footmisc}

\usepackage[T1]{fontenc}
\usepackage[utf8]{inputenc}
\usepackage{lmodern}

\usepackage[colorlinks, allcolors=blue!70!black, linktocpage]{hyperref}

\setlist[enumerate]{noitemsep, label=(\arabic*), ref=(\arabic*)}

\numberwithin{equation}{section}

\usepackage{cleveref}

\usepackage{microtype}

\usepackage{floatrow}
\let\OLDtableofcontents\tableofcontents
\renewcommand\tableofcontents[1]{%
    {\baselineskip 0.5ex %
	\OLDtableofcontents{#1}}%
}

\let\OLDthebibliography\thebibliography
\renewcommand\thebibliography[1]{%
	\setstretch{1.079} 
	\OLDthebibliography{#1}%
	\small %
	\setlength{\itemsep}{0.2\baselineskip} 
}

\let\OLDfootnote\footnote
\renewcommand\footnote[1]{%
	\setlength{\footnotesep}{0.75\baselineskip}%
	{\footnotesize \OLDfootnote{#1}}%
}

\theoremstyle{remark}
\newtheorem{remark}{Remark}[section]

\crefname{equation}{Eq.}{Eqs.}
\creflabelformat{equation}{#2#1#3}

\crefname{section}{Sec.}{Sec.}
\crefname{appendix}{Appendix}{Appendices}
\crefname{figure}{Fig.}{Figs.}

\crefname{definition}{Def.}{Defs.}
\crefname{prop}{Prop.}{Props.}
\crefname{lemma}{Lemma}{Lemmas}
\crefname{corollary}{Cor.}{Cors.}
\crefname{thm}{Theorem}{Theorems}
\crefname{remark}{Remark}{Remarks}

\renewcommand\thesection{\arabic{section}}
\renewcommand\thesubsection{\arabic{subsection}}

\makeatletter
\def\p@subsection{\thesection.}
\def\p@subsubsection{\thesection.\thesubsection.}
\makeatother 

\newcommand{\lie}{\pounds}

\newcommand{\mc}{\mathcal}

\newcommand{\ms}{\mathscr}
\newcommand{\mf}{\mathfrak}
\newcommand{\bb}{\mathbb}

\newcommand{\df}[1]{\boldsymbol{#1}}

\newcommand{\hateq}{\mathrel{\mathop {\widehat=} }} 

\newcommand{\eqsp}{\, ,\quad} 

\newcommand{\lb}{\left}
\newcommand{\rb}{\right}

\let\oldint\int
\renewcommand{\int}{\oldint\limits}

\renewcommand{\bar}{\overline}

\newcommand{\Hajicek}{H\'a\'{\j}i\v{c}ek}

\def\be{\begin{equation}}
\def\ee{\end{equation}}

\begin{document}

\setstretch{1.2}


\title{Symmetries, charges and conservation laws at causal diamonds in general relativity}

\author{Venkatesa Chandrasekaran}\email{ven\_chandrasekaran@berkeley.edu}
\affiliation{Center for Theoretical Physics and Department of Physics, University of California, Berkeley, CA 94720}

\author{Kartik Prabhu}\email{kartikprabhu@cornell.edu}
\affiliation{Cornell Laboratory for Accelerator-based Sciences and Education (CLASSE), Cornell University, Ithaca, NY 14853}

\begin{abstract}
We study the covariant phase space of vacuum general relativity at the null boundary of causal diamonds. The past and future components of such a null boundary each have an infinite-dimensional symmetry algebra consisting of diffeomorphisms of the \(2\)-sphere and boost supertranslations corresponding to angle-dependent rescalings of affine parameter along the null generators. Associated to these symmetries are charges and fluxes obtained from the covariant phase space formalism using the prescription of Wald and Zoupas. By analyzing the behavior of the spacetime metric near the corners of the causal diamond, we show that the fluxes are also Hamiltonian generators of the symmetries on phase space. In particular, the supertranslation fluxes yield an infinite family of boost Hamiltonians acting on the gravitational data of causal diamonds. We show that the smoothness of the vector fields representing such symmetries at the bifurcation edge of the causal diamond implies suitable matching conditions between the symmetries on the past and future components of the null boundary. Similarly, the smoothness of the spacetime metric implies that the fluxes of all such symmetries are conserved between the past and future components of the null boundary. This establishes an infinite set of conservation laws for finite subregions in gravity analogous to those at null infinity. We also show that the symmetry algebra at the causal diamond has a non-trivial center corresponding to constant boosts. The central charges associated to these constant boosts are proportional to the area of the bifurcation edge, for any causal diamond, in analogy with the Wald entropy formula.
\end{abstract}

\maketitle 
\tableofcontents

\section{Introduction}
\label{sec:intro}

Understanding the description of finite subsystems in diffeomorphism invariant theories is an important problem in both classical and quantum gravity. Over the years there has been a lot of progress in uncovering various aspects of gravitational subsystems by studying the covariant phase space formalism in the presence of boundaries \cite{LW,IW-noether-entropy, WZ,Carlip:1998wz,Balachandran:1994up, Donnelly:2016auv, Speranza:2017gxd,Kirklin:2019xug, Strominger:2017zoo, Harlow:2019yfa}. The presence of a boundary promotes a subset of the boundary preserving diffeomorphisms to symmetries of the covariant phase space. These boundary symmetries then result in non-trivial boundary charges which can be thought of as capturing aspects of the degrees of freedom contained within the subregion.

Null boundaries are particularly important as they play a fundamental role in gravitational thermodynamics \cite{Wall:2011hj, Padmanabhan:2013nxa, Hopfmuller:2018fni,Jacobson:1995ab}, as well as in holography and quantum gravity \cite{Wieland:2017zkf, Bousso:1999xy, Wieland:2019hkz}. Moreover, it was recently conjectured that the symmetries and charges at stationary event horizons are relevant to the black hole information problem \cite{Hawking:2016msc, Hawking:2016sgy, Haco:2018ske}. A particularly important class of null surfaces are the boundaries of causal diamonds, which are fundamental to the description of gravitational subregions. Black hole thermodynamics \cite{IW-noether-entropy,Bardeen:1973gs,PhysRevD.7.2333,Wall:2011hj} and entanglement entropy in AdS/CFT \cite{Ryu:2006bv} have demonstrated that geometric properties of causal horizons are deeply related to the thermodynamic and statistical properties of spacetime. This strongly suggests that these deep connections generalize to any causal diamond in any spacetime. While there have been important insights in this direction \cite{Jacobson:2015hqa, Jacobson:2018ahi,deBoer:2016pqk}, a complete understanding for arbitrary gravitational subregions remains elusive. A potential avenue of progress lies in the covariant phase space formalism applied to gravity at the boundaries of causal diamonds. 

In \cite{CFP} the covariant phase space of general relativity at null boundaries was studied in detail. There it was shown that there exists an infinite-dimensional symmetry algebra for general relativity at \emph{all} null boundaries, including non-stationary event horizons, in any spacetime. For null boundaries of the form $N = \mathcal{Z}\times\mathbb{R}$, where \(\mc Z\) is the space of null generators, the algebra takes the form $\text{diff}(\mathcal{Z})\ltimes\mathfrak{s}$ with the ``supertranslation'' subalgebra $\mathfrak{s}$ consisting of angle-dependent translations and rescalings of affine parameter along the null surface. The charges (at cross-sections of the null surface) and fluxes associated to the symmetries were computed from the covariant phase space formalism using the prescription given by Wald and Zoupas \cite{WZ}.

In this paper we use the results of \cite{CFP} at the boundaries of causal diamonds. We consider causal diamonds obtained from the intersection of the chronological past and future of timelike separated points in a convex normal neighborhood. The boundary of such a causal diamond is a null surface $N$ with a \(2\)-sphere bifurcation edge $B$. We use Gaussian null coordinates adapted to the causal diamond boundary to show that both $N$ and $B$ for any causal diamond in any spacetime can be identified. The resulting reduced symmetry algebra which preserves \(B\) takes the form \(\rm{diff(\bb S^2}) \ltimes \mf b\) with \(\mf b\) consisting of the angle-dependent rescalings along the null generators. The angle-dependent translations along the null generators are eliminated by requiring the bifurcation edge \(B\) to be preserved.

By considering the behavior of geometric fields on $N$ near its corners, we show that the boundary charges and the boundary presymplectic potential vanish in the limit to the corners of the causal diamond. From this we show that the Wald-Zoupas fluxes are Hamiltonian generators on the covariant phase space, which in particular provides an infinite family of boost generators for any smooth causal diamond in general relativity. This is similar in spirit to the boost generator at Killing horizons, which is also the vacuum modular Hamiltonian, where in the present context the boost generators act on the gravitational data associated to the causal diamond. Furthermore, we show that the reduced symmetry algebra at $N$ has a non-trivial center. The charges associated to the elements of the center are precisely the boost generators, whose values are proportional to the area of $B$. Thus there exists a Wald entropy \cite{IW-noether-entropy}, and a quasi-local first law, for any smooth causal diamond in general relativity. 

Using the smoothness of the spacetime metric and the vector fields representing the null boundary symmetries we then show that the Wald-Zoupas fluxes associated to the symmetries are conserved between the past and future components of $N$. This gives an infinite set of conservation laws for finite subregions in general relativity on any spacetime. This is analogous to the conservation laws between past and future null infinity \cite{Strominger:2013jfa, Tro, Prabhu:2018gzs, Prabhu:2019fsp} except, in this case, the smoothness of fields at $B$ are much simpler to analyze. Just as the asymptotic conservation laws between past and future null infinity place an infinite number of constraints on gravitational scattering (conjectured to hold even in the quantum theory \cite{Strominger:2013jfa}), the conservation laws we derive for finite causal diamonds likely place important constraints on the properties of scattering in local gravitational subsystems.

The rest of the paper is organized as follows. In \cref{sec:review} we review the formulation of symmetries and the associated charges and fluxes in general relativity from \cite{CFP}, and reformulate them in terms of Gaussian null coordinates. In \cref{sec:causal-diamond} we adapt this formalism to the boundaries of causal diamonds and detail the reduction of the symmetry algebra of a general null surface to a subalgebra which preserves the structure on a causal diamond. We also investigate the behavior of the fields and charges near the corners of the causal diamond and show that the fluxes associated to the symmetries at a causal diamond are also Hamiltonian generators on the corresponding phase space. In \cref{sec:conservation} we show that the smoothness of the spacetime metric at the causal diamond implies an infinite number of conservation laws for the fluxes through the null boundaries. In \cref{sec:central-charge} we compute the charges associated to the central elements of the symmetry algebra and show that these take the form of a ``first law''. We end with \cref{sec:disc} summarizing and discussing the potential applications of our results. In \cref{sec:WZ} we collect the essential ingredients of the covariant phase space formalism and the Wald-Zoupas prescription for calculating charges and fluxes. In \cref{sec:central-ext} we analyze the structure of the symmetry algebra at a causal diamond and show how it arises as a non-trivial central extension.

\subsection*{Notation and conventions}
We follow the conventions of \cite{Wald-book}. We use abstract indices \(a,b, \ldots\) to denote tensor fields, e.g. \(g_{ab}\) is the spacetime metric, and indices \(A,B, \ldots\) to denote components of tensor fields in some coordinate system on \(\bb S^2\), e.g. \(q_{AB}\) is a metric on \(\bb S^2\). Boldface quantities like $\df \omega$ will denote differential forms. 

We also use the following terminology for the charges associated to the symmetry algebra at a null boundary \(N\). Quantities associated to null boundary symmetries evaluated as integrals over cross-sections of the null boundaries will be called ``charges'', while the difference of these charges on two cros-sections evaluated as an integral over a portion of the null boundary will be called ``fluxes''. When certain conditions are satisfied the fluxes can also be considered as Hamiltonian generators on the null boundary phase space (see \cref{eq:gen-cond0,eq:gen-cond}).

\section{Null boundary symmetries and charges} \label{sec:review}

In this section we briefly review the basic formalism and results of \cite{CFP}, namely the symmetries and charges at a null boundary in general relativity. We then recast the null boundary phase space, and the resulting symmetries and charges, in terms of Gaussian null coordinates. This will prove to be useful when considering causal diamonds.

The relationship between the covariant approach of \cite{CFP} which is intrinsic to the null boundary and the coordinate-based approach in \cref{sec:GNC,sec:symm-GNC} is the same as that between the intrinsic universal structure approach \cite{Geroch-asymp,Ashtekar:1981bq,Ashtekar:2014zsa} and one based on Bondi coordinates \cite{Bondi:1962px,Sachs:1962wk,Sachs:1962zza} or the conformal Gaussian null coordinates \cite{Hol-Th,Hollands:2016oma} at null infinity in asymptotically-flat spacetimes.

\subsection{Universal structure and symmetries on null boundaries}
\label{sec:CFP-summ}

Consider a spacetime $(M, g_{ab})$ with null boundary $N$. For now we will assume the null generators of $N$ are complete, i.e. $N \cong \mathcal{Z}\times \mathbb{R}$, where $\mathcal{Z}$ is the space of null generators. Later we will consider null surfaces with boundary, which is the relevant setting for causal diamonds. The null boundary \(N\) is naturally equipped with the equivalence class $[\ell^a, \kappa]$ where $\ell^a$ is the null generator of $N$, $\kappa$ is the non-affinity defined by\footnote{We use the notation $\hateq$ to mean `equality on $N$' throughout the paper.} 
\begin{align}
\ell^b \nabla_b \ell^a \hateq \kappa \ell^a,
\end{align}
and the equivalence class $[\ell^a, \kappa]$ is defined by the rescaling freedom
\be\label{eq:rescaling}
\ell^a \mapsto e^{\beta}\ell^a  \eqsp \kappa \mapsto e^{\beta}(\kappa + \lie_{\ell}\beta)
\ee
where $\beta$ is a smooth function on $N$. 

In \cite{CFP} it was shown that the structure $[\ell^a, \kappa]$ is \emph{universal} in the sense that different such structures on $N$, as induced by different background metrics, are all related by diffeomorphisms (we shall show this explicitly in \cref{sec:GNC} in a Gaussian null coordinate system). We can then define the field configuration space $\mathscr{F}$ to be the set of smooth metrics $g_{ab}$ on a manifold $M$ with null boundary $N$ which is equipped with the universal structure $[\ell^a, \kappa]$. The covariant phase space is the subset $\bar{\mathscr{F}} \subset \mathscr{F}$ consisting of on-shell metrics satisfying the vacuum Einstein equation.\\

The group of symmetries on a null boundary \(N\) is the subgroup of diffeomorphisms on $M$ which preserves the null boundary $N$ and the universal structure on it. It will be easier to work with the symmetry algebra instead of the group. The symmetry algebra on the null boundary consists of vector fields $\xi^a$ on $M$ which are tangent to \(N\) and preserve the linearized version of the equivalence relation \cref{eq:rescaling}. This results in the conditions
\be\label{symmetryalgebra}\begin{aligned}
\lie_{\xi}\ell^a &\hateq \beta \ell^a \\
\lie_{\xi}\kappa &\hateq \beta \kappa + \lie_{\ell}\beta, 
\end{aligned}\ee
where \(\beta\) is some smooth function on \(N\) which depends on the vector field \(\xi^a\). The detailed structure of the resulting symmetry algebra \(\mf g\) was derived in \cite{CFP} and can be summarized as follows. The vector fields of the form \(\xi^a \hateq f \ell^a\) with \(\lie_{\ell}(\lie_{\ell} + \kappa )f \hateq 0\) form an infinite-dimensional abelian Lie ideal \(\mf s \subset \mf g\) of \emph{supertranslations}. The quotient algebra $\mathfrak{g}/\mathfrak{s}$ is isomorphic to $\text{diff}(\mathcal{Z})$, the algebra of smooth diffeomorphisms of the space of null generators \(\mc Z\). There is an additional Lie ideal \(\mf s_0 \subset \mf s\) of \emph{affine supertranslations} given by \(\xi^a \hateq f \ell^a\) with \((\lie_\ell + \kappa)f \hateq 0\). Hence the symmetry algebra \(\mf g\) can be written as 
\begin{align}\label{eq:g-semidirect}
\mathfrak{g} \cong \text{diff}(\mathcal{Z})\ltimes (\mf b \ltimes \mf s_0)
\end{align}
where \(\mf b \cong \mf s/ \mf s_0\).

The charges and fluxes associated to this symmetry algebra were also derived in \cite{CFP} using the Wald-Zoupas procedure. Writing the covariant expression for these charges would require introducing a significant amount of formalism and notation. Instead we will derive the symmetry algebra, and express the charges and fluxes, using Gaussian null coordinates in \cref{sec:symm-GNC}.

\subsection{Gaussian null coordinates}
\label{sec:GNC}

It will be convenient to introduce coordinates adapted to the null surface \(N\) called \emph{Gaussian null coordinates} (GNC) \cite{Penrose}, which have been used in a variety of contexts \cite{Donnay:2015abr, Hollands:2016oma, Morales:2008,Parattu:2015gga,Madler}. We briefly review the construction of GNC below. Since our main interest is in causal diamonds, we will now restrict (for convenience) to the case of \(4\)-dimensional spacetimes where the space of null generators is a \(2\)-sphere \(\mc Z = \bb S^2\), but our results can be readily generalized.

Let \(\ell^a\) be an affinely-parameterised (\(\kappa = 0\)) null normal to \(N\) and let \(v\) be an affine parameter along these null generators i.e., \(v\) is some function on \(N\) such that \(\ell^a \nabla_a v \hateq 1\). Now let \(S \cong \bb S^2\) be a cross-section of \(N\) such that \(v\vert_S = 0\), and let \(x^A\) be a coordinate system on \(S\). We extend the coordinate functions \(x^A\) to all of \(N\) by parallel-transport along the null generators, \(\ell^a \nabla_a x^A \hateq 0\). This defines a coordinate system \((v, x^A)\) on \(N\).

To define a coordinate system in a neighborhood of \(N\), let \(u\) be a function in such a neighborhood so that \(u\vert_N = 0\) on \(N\). Then, \(\ell_a \equiv - du\) is the normal to \(N\) and the vector field \(n^a \equiv \partial/\partial u\) is transverse (i.e. not tangent) to \(N\). To fix coordinates away from \(N\) we choose \(u\) such that \(n^a\) is an affinely-parameterised null vector field i.e. \(n^a n_a = 0\)  and \(n^b\nabla_b n^a = 0\). Then we extend the coordinates \((v,x^A)\) away from \(N\) by parallel transport along \(n^a\). The coordinate functions \((u,v, x^A)\) define a GNC in a neighborhood of the null surface \(N\). It follows from the above definition of the GNC that in these coordinates the spacetime metric satisfies \cite{Madler}
\be\label{gauge}\begin{aligned}
g_{uu} &= g_{uA} = 0 \eqsp g_{uv} = -1 \\ 
g_{vv} &\hateq g_{vA} \hateq \partial_u g_{vv} \hateq 0 
\end{aligned}\ee
and thus we can write the line element in the form (this is equivalent to the form in \cite{Penrose})
\be\label{eq:g-GNC}\begin{aligned}
    & ds^2 = -W dv^2 -2du dv + q_{AB}(dx^A - W^A dv)(dx^B - W^B dv) \\
    \text{where} \quad& W\vert_{u = 0} = \partial_{u}W\vert_{u = 0} = W^A\vert_{u = 0} = 0
\end{aligned}\ee
and $W$, $W^A$, $q_{AB}$ are functions of $(u,v,x^A)$, and can be considered as tensors on \(\bb S^2\) which depend on \((u,v)\). The tensor \(q_{AB}\) defines a Riemannian metric on the \(2\)-spheres of constant \(u\) and \(v\). The extensions of the null generator $\ell^a$ and the auxilliary null vector $n^a$ in the neighborhood of \(N\) are given by 
\be
    \ell^a \equiv \partial_v -\tfrac{1}{2}W \partial_{u} + W^A\partial_A \eqsp n^a \equiv \partial_u 
\ee 

The shear \(\sigma_{AB}\) and expansion \(\theta\) of \(N\) are given by the relation
\be\label{eq:shear-exp-GNC}
    \tfrac{1}{2}\partial_v q_{AB} \hateq \sigma_{AB} + \tfrac{1}{2} q_{AB} \theta
\ee
where \(\sigma_{AB} q^{AB} = 0\), while the \Hajicek\ rotation \(1\)-form of the \(u = \text{constant}\) cross-sections is given by
\be\label{eq:rotation-GNC}
    \omega_A \hateq - q_A{}^a n_b \nabla_a \ell^b  = - \tfrac{1}{2} \partial_u (q_{AB}W^B)
\ee

We emphasize that the above construction of the GNC can be carried out in any spacetime in a neighborhood of any null surface. Now let \((M_1,g_1)\) and \((M_2,g_2)\) be two spacetimes with null surfaces \(N_1\) and \(N_2\) along with the GNCs \((u_1,v_1,x^A_1)\) and \((u_2,v_2,x^A_2)\), as constructed above, respectively. Without any loss of generality we can identify a neighborhood of \(N_1\) in \((M_1,g_1)\) with that of \(N_2\) in \((M_2,g_2)\) by identifying the corresponding GNCs \((u_1,v_1, x^A_1) = (u_2, v_2, x^A_2)\). Thus, we can identify all the spacetimes under consideration, and work on a single manifold \(M\) with boundary \(N\) such that the configuration space \(\ms F\) consists of all the metrics in \(M\) for which \(N\) is a null surface and the metrics take the form \cref{eq:g-GNC} in a GNC in a neighborhood of \(N\). From the above construction we see that while the induced metric \(q_{AB}\) depends on the particular choice of the spacetime metric in \(\ms F\), the null generator \(\ell^a\) is common to all metrics in \(\ms F\). Thus, the null surface \(N\) along with the affine null generator \(\ell^a\) are \emph{universal}. Note that we can construct the GNC even with a non-affinely parametrized \(\ell^a\), which leads to the universal structure used in \cite{CFP} as described in \cref{sec:CFP-summ}.

\subsection{Null boundary symmetries, charges and fluxes in GNC}
\label{sec:symm-GNC}

The symmetry algebra at the null boundary \(N\) consists of the vector fields $\xi^a$ generating infinitesimal coordinate transformations which preserve the GNC form of the metric in \cref{eq:g-GNC}. We expand the vector field $\xi^a$ in the GNC to first order in $u$ as, 
\begin{align}
\xi^a \equiv (f_0 + u f_1) \partial_v + (\beta_0 + u\beta_1) \partial_u + (X^A_0 + u X^A_1)\partial_A + O(u^2)
\end{align}
To preserve the location $u = 0$ of the null surface $N$, \(\xi^a\) must be tangent to \(N\) and hence $\beta_0 = 0$. To preserve the form of the metric \cref{eq:g-GNC} we have
\begin{subequations}
\begin{align}
	\lie_\xi g_{uu} &= \lie_{\xi} g_{uv} = \lie_{\xi}g_{uA} = 0 \label{cond1}\\
	\lie_\xi g_{vA} &= O(u) \label{cond2}  \\
	\lie_\xi g_{vv} &= O(u^2) \label{cond3}
\end{align}
\end{subequations} 
Evaluating \cref{cond1} at $u = 0$ we have
\begin{subequations}\begin{align}
	\lie_\xi g_{uu} = 0 \implies& f_1 = 0 \\
    \lie_\xi g_{uv} = 0 \implies& \beta_1 = -\partial_v f_0 \label{eq:beta1-f0} \\
    \lie_\xi g_{uA} = 0 \implies& q_{AB}X^A_1 = \partial_A f_0 
\end{align}\end{subequations}
The conditions \cref{cond2} imply
\be\label{eq:X-cond}
    \partial_v X^A_0 = 0
\ee
while \cref{cond3}, evaluated to $O(u)$ using \cref{eq:X-cond}, gives
\be\label{eq:beta1-cond}
    \partial_v \beta_1 = 0 \implies \partial_v^2 f_0 = 0
\ee
where the second condition follows from \cref{eq:beta1-f0}. Similar conditions were derived independently in \cite{Madler}.

From \cref{eq:X-cond,eq:beta1-cond} we conclude that the symmetries on the null boundary \(N\) are characterized by \((\alpha,\beta, X^A)\) where \(\alpha\) and \(\beta\) are functions and \(X^A\) is a vector field on \(\bb S^2\). In a neighborhood of \(N\) this symmetry is represented by a vector field \(\xi^a\), which in GNC takes the form
\be\label{eq:symm-GNC}
	\xi^a \equiv (\alpha - v \beta) \partial_v + u \beta \partial_u + \lb[ X^A + u q^{AB}\partial_B (\alpha - v \beta) \rb] \partial_A + O(u^2)
\ee 
Note that the vector field \(\xi^a\) at \(N\), i.e. \(u = 0\), is parametrized entirely by \((\alpha, \beta, X^A)\) and is independent of the choice of metric in the configuration space \(\ms F\).

We now analyze the structure of the symmetry algebra \(\mf g\) generated by such vector fields. Consider two symmetries \(\xi_1 = (\alpha_1, \beta_1,X^A_1)\) and \(\xi_2 = (\alpha_2, \beta_2,X^A_2)\) in \(\mf g\). Their Lie bracket can be computed using their representations in terms of vector fields as in \cref{eq:symm-GNC}. A straightforward computation gives 
\begin{subequations}\label{eq:bracket}\begin{align}
    &\lb[ (\alpha_1, \beta_1, X^A_1), (\alpha_2, \beta_2, X^A_2) \rb] = (\alpha, \beta, X^A) \\
    \text{where} \quad 
    \alpha &= - \alpha_1 \beta_2 + \alpha_2 \beta_1 + X^A_1 \partial_A \alpha_2 - X^A_2 \partial_A \alpha_1 \\
    \beta &= X^A_1 \partial_A \beta_2 - X^A_2 \partial_A \beta_1 \label{eq:bracket-beta}\\
    X^A &= \lb[ X_1, X_2 \rb]^A = X^B_1 \partial_B X^A_2 - X^B_2 \partial_B X^A_1
\end{align}\end{subequations}
where the last line is the Lie bracket of vector fields on \(\bb S^2\). Note that the sign of \(\beta\) in Eq.~4.10b \cite{CFP} is incorrect and has been corrected in \cref{eq:bracket-beta} above.

From \cref{eq:bracket} it is easy to deduce the following structure of \(\mf g\). If \(X^A_1 = 0\) then \(X^A = 0\), i.e. symmetries of the form \((\alpha,\beta, 0)\) form an abelian Lie ideal \(\mf s \subset \mf g\) of \emph{supertranslations}. The quotient \(\mf g/ \mf s\) is then isomorphic to the Lie algebra \({\rm diff}(\bb S^2)\) represented by symmetries of the form \((0,0, X^A)\). There is an additional Lie ideal in \(\mf g\) which is given as follows. In \cref{eq:bracket}, taking \(\beta_1 = X^A_1 = 0\) we get \(\beta = X^A = 0\), that is, symmetries of the form \((\alpha,0,0)\) are also an abelian Lie ideal \(\mf s_0 \subset \mf g\) called \emph{affine supertranslations}. The quotient \(\mf b \cong \mf s/\mf s_0\) of all the supertranslations by \(\mf s_0\) is represented by symmetries of the form \((0,\beta,0)\). Thus, the symmetry algebra \(\mf g\) on any null boundary has the structure (same as in \cref{eq:g-semidirect})
\be
    \mf g \cong {\rm diff}(\bb S^2) \ltimes (\mf b \ltimes \mf s_0)
\ee
It was shown in \cite{CFP} that this symmetry algebra coincides with the definition given by Wald and Zoupas \cite{WZ}, reviewed below \cref{eq:omega-Q-int}.

\begin{remark}[Symmetry group at \(N\)]\label{rem:group}
    The symmetry group can also be obtained by considering finite coordinate transformations of the GNC which preserve the metric form \cref{eq:g-GNC}. In particular at \(N\), i.e. \(u = 0\), we have the coordinate transformations \((v, x^A) \mapsto (\bar v, \bar x^A)\) with
\be
    \bar v = \alpha(x^A) + e^{-\beta(x^A)}v \eqsp \bar x^A = \bar x^A(x^B) 
\ee
Thus, the symmetry group consists of arbitrary diffeomorphisms of \(\bb S^2\) along with angle-dependent translations (given by \(\alpha(x^A)\)) and angle-dependent rescalings (given by \(\beta(x^A)\)) along the null generators.
\end{remark}

When the null surface \(N\) has additional structure which is also universal --- i.e., common to \emph{all} the spacetimes under consideration --- the symmetry algebra can be reduced further. For instance, when all the spacetimes have a Killing vector field in a neighborhood of \(N\) which becomes tangent to \(N\), the symmetries proportional to this Killing field provide a preferred \(1\)-dimensional subalgebra of \(\mf g\) (see Sec.~4.4 \cite{CFP}). Similarly, if the null surface is stationary for all spacetimes, so that the shear and expansion of \(N\) vanish, then the symmetry algebra can be reduced so that \(\beta = \text{constant}\) and \(X^A\) is a conformal Killing field on \(\bb S^2\) i.e., an element of the Lorentz algebra (see Sec.~IV.B \cite{Ash-Bah}).\footnote{Note that the reduction of \({\rm diff}(\bb S^2)\) to the Lorentz algebra in the stationary case relies crucially on the \(\bb S^2\) topology of the cross-sections.} We show in \cref{sec:symm-CD} that when \(N\) is the null boundary of causal diamonds a similar reduction of the symmetry algebra occurs due to the presence of a preferred cross-section corresponding to the bifurcation edge. Specifically, since the bifurcation edge is a preferred cross-section of the null boundary of a causal diamond, only those symmetry vector fields which preserve its location i.e. are tangent to the bifurcation surface are permitted in the symmetry algebra. This has the effect of eliminating the affine supertranslations \(\xi^a \equiv \alpha \partial_v \in \mf s_0\) from the symmetry algebra (see also Sec.~4.5 \cite{CFP}).\\

The charges and fluxes associated to the null boundary algebra \(\mf g\) were computed in \cite{CFP} using the covariant phase space formalism along with the Wald-Zoupas prescription. It was also shown that the ambiguities in the symplectic current and the Wald-Zoupas prescription do not affect the resulting charges and fluxes. We do not repeat the full analysis of \cite{CFP}, but below we write down the relevant expressions for the boundary presymplectic potential $\df{\Theta}(g;\delta g)$, the Wald-Zoupas (WZ) charges $\mc Q_{\xi}$ and fluxes $\mc F_{\xi}$ for vacuum general relativity, derived in \cite{CFP}, in terms of GNC.

The boundary presymplectic potential on \(N\) is given by
\begin{align}\label{WZcorrection}
\df{\Theta}(g; \delta g) = \frac{1}{16\pi}\df{\varepsilon}_3 \left(\sigma^{AB} - \tfrac{1}{2}q^{AB}\theta\right) q_A{}^a q_B{}^b \delta g_{ab}
\end{align}
where \(\df\varepsilon_3 \equiv \varepsilon_{abc}\) is the \(3\)-volume element on \(N\).

Let \(S\) be any cross-section of \(N\) with area-element \(\df\varepsilon_2 \equiv \varepsilon_{ab}\) and \(\Delta N\) be a region of \(N\) bounded by two cross-sections. The charges (on \(S\)) and fluxes (through \(\Delta N\)) associated to a supertranslation $\xi^a \hateq f\ell^a = (\alpha - v\beta) \ell^a$ are
\be\label{supcharge}\begin{aligned}
    \mathcal{Q}_{f}[S] &= \frac{1}{8\pi}\int_{S}\df{\varepsilon}_2 \left[ (\alpha - v \beta) \theta + \beta \right] \\
    \mc F_{f}[\Delta N] &= \frac{1}{8\pi}\int_{\Delta N}\df{\varepsilon}_3~ (\alpha - v \beta) \left(\sigma_{AB} \sigma^{AB} - \tfrac{1}{2}\theta^2 \right)
\end{aligned}\ee
while those associated to a $\text{diff}(\bb S^2)$ generator $X^A$ (taken to be tangent everywhere to the \(v = \text{constant}\) cross-sections of \(N\)) are given by 
\be\label{diffcharge}\begin{aligned}
    \mathcal{Q}_{X}[S] &= \frac{1}{8\pi}\int_{S}\df{\varepsilon}_2 \left( - \omega_A X^A \right) \\
    \mc F_X[\Delta N] &= \frac{1}{8\pi}\int_{\Delta N}\df{\varepsilon}_3 \left(\sigma_{AB} - \tfrac{1}{2} \theta q_{AB} \right) D^A X^B
\end{aligned}\ee
where the \Hajicek\ rotation \(1\)-form \(\omega_A\) (in the GNC foliation) is as defined in \cref{eq:rotation-GNC}, and \(D_A\) is the derivative operator with respect to \(q_{AB}\) in the foliation given by the GNC. 

Note that the supertranslation charge expression \cref{supcharge} can be evaluated on any choice of cross-section, while the charge expression \cref{diffcharge} only holds on the cross-sections of the \(v = \text{constant}\) foliation.\footnote{The dependence of the \({\rm diff}(\bb S^2)\) charge expression on the foliation is a result of the semidirect structure of \(\mf g\) in \cref{eq:g-GNC}, that is, there does not exist any unique choice of a \({\rm diff}(\bb S^2)\) subalgebra of \(\mf g\). This is similar to the status of the Lorentz algebra within the BMS algebra at null infinity.} If the \({\rm diff}(\bb S^2)\) charge is to be evaluated on some arbitrary foliation of \(N\) with normal \(\hat n_a\) such that \(\ell^a \hat n_a \hateq -1\), then we have instead
\be\label{diffcharge-alt}
    \mathcal{Q}_{\hat X}[\hat S] = \frac{1}{8\pi}\int_{\hat S}\hat{\df\varepsilon}_2 \left( \beta_{\hat X} - \hat \omega_A \hat X^A \right)
\ee
where now \(\hat X^A\) is taken to be tangent to the cross-sections of the chosen foliation, \(\hat \omega_A\) is the corresponding \Hajicek\ rotation \(1\)-form, while \(\beta_{\hat X}\) is given by
\be\label{eq:betaX-defn}
    \beta_{\hat X} = - \hat X^A \hat q_A{}^a \lie_\ell \hat n_a
\ee

\begin{remark}[Fluxes from the charges]
    Given the charge expressions \cref{supcharge,diffcharge}, the corresponding fluxes can be obtained using the vacuum Einstein equations \(R_{ab} = 0\) on \(N\). Specifically we have
    \be\label{eq:constraints}\begin{aligned}
    R_{ab}\ell^a \ell^b &= 0 \implies \partial_v \theta &&= -\tfrac{1}{2} \theta^2 - \sigma_{AB} \sigma^{AB} \\
    R_{ab}q_A{}^a \ell^b &= 0 \implies  \partial_v \omega_A && = - \theta \omega_A - D^B \sigma_{AB} + \tfrac{1}{2} D_A \theta
    \end{aligned}\ee
    where \(D_A\) is the derivative operator with respect to \(q_{AB}\) in the foliation given by the GNC. These are the \emph{Raychaudhuri equation} and the \emph{Damour-Navier-Stokes equation} respectively \cite{Gourgoulhon:2005ng, Booth:2012xm}.
\end{remark}

\section{Causal Diamonds}
\label{sec:causal-diamond}

In this section we apply the above results to the boundaries of causal diamonds, with the appropriate modifications necessary for null surfaces with boundary. We begin by recalling the definition of causal diamonds and the structure on their null boundaries.

\begin{figure}[h!]
    \centering
    \includegraphics[width=0.45\textwidth]{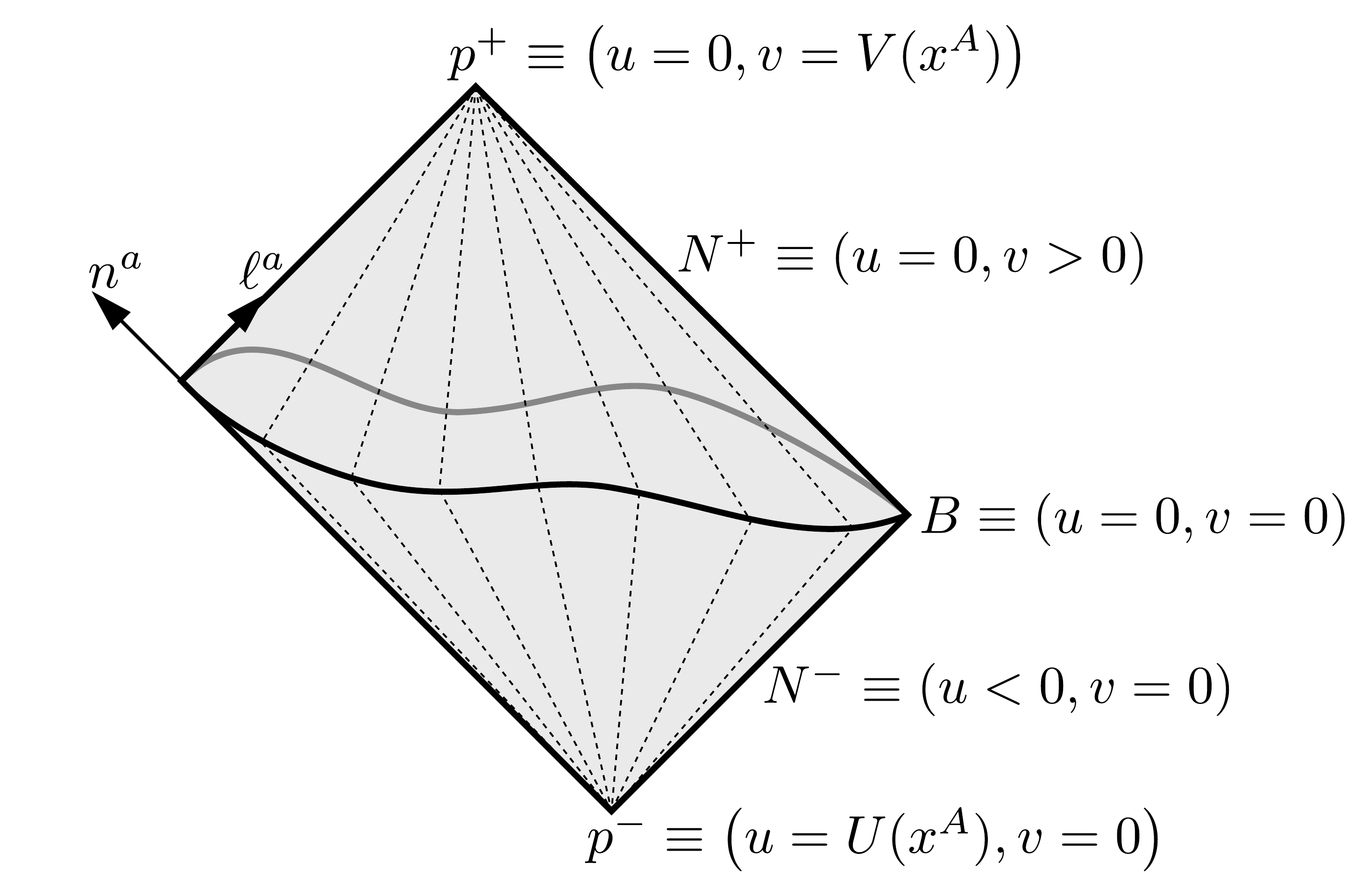}
    \caption{Diagram of a causal diamond in a spacetime \((M,g)\). The points \(p^\pm\) denote the corners of the causal diamond and \(B\) is the bifurcation edge while \(N^\pm\) denote the future/past null surfaces joining \(B\) to \(p^\pm\), respectively. The functions \(v\) and \(u\) are affine coordinates with affine null normals \(\ell^a\) and \(n^a\) on \(N^\pm\).}
    \label{fig:causal-diamond}
\end{figure}

In a given spacetime \((M,g)\), consider two points $p^+, p^- \in M$ such that $p^+$ is in a convex normal neighborhood of $p^-$ and is in its chronological future, i.e., \(p^+\) is ``inside the future light cone'' of \(p^-\). The intersection of the chronological past of $p^+$ with the chronological future of $p^-$ defines a \emph{causal diamond} or a \emph{double cone} (see for instance \cite{Witten:2019qhl,DPP,Jacobson:2018ahi,deBoer:2016pqk}). We assume that the causal diamond is ``small enough'' so that the null generators emanating from \(p^\pm\) form smooth null surfaces \(N^\pm\) respectively, which intersect at a smooth \(2\)-surface \(B\), the \emph{bifurcation edge}, which is topologically \(\bb S^2\). We denote the null boundary of the causal diamond by \(N = N^+ \cup N^-\) (see \cref{fig:causal-diamond}).

We now investigate the null boundary symmetries and charges defined in \cite{CFP} for the null boundary \(N\) of a causal diamond. It will be convenient to use the formulation in terms of GNC as detailed in \cref{sec:GNC}, which we adapt to a causal diamond as follows.

Unlike a general null surface, the boundary of a causal diamond has a preferred cross-section determined by the bifurcation edge \(B\). At \(B\) there exist \emph{unique} null vector fields \(\ell^a\) and \(n^a\) both future-directed such that \(\ell^a\) is tangent to \(N^+\), \(n^a\) is tangent to \(N^-\), and \(\ell^a n_a\vert_B = -1\). We can extend these vector fields to \(N^\pm\) so that \(\ell^a\) is the affine null generator of \(N^+\), \(n^a\) is the affine null generator of \(N^-\), and \(\ell^a n_a\vert_N = -1\). Let \(x^A\) be some coordinates on \(B\); we pick the affine parameter \(v\) of \(\ell^a \equiv \partial/\partial v\) on \(N^+\), and similarly \(u\) of \(n^a \equiv \partial/\partial u\) on \(N^-\), such that \(B \equiv (u=0,v=0)\). Since, \(\ell^a\) is future-directed, the coordinate \(v\) increases moving towards \(p^+\) from \(v= 0\) at \(B\). Note that in a general spacetime \(N^+\) will have both a shear and an expansion, which depend on the space of generators, hence the value of the affine-parameter \(v\) at the corner \(p^+\) will depend on the null generator along which we approach \(p^+\) i.e. \(v\vert_{p^+} = V(x^A)\). Similarly, on \(N^-\) the affine parameter \(u\) along \(n^a\) decreases moving towards \(p^-\) from \(u= 0\) at \(B\) and \(p^-\) lies at \(u\vert_{p^-} = U(x^A)\). These are depicted in \cref{fig:causal-diamond}.

As described in \cref{sec:GNC}, we can extend \((u,v,x^A)\) to form a GNC in a neighborhood of the causal diamond.\footnote{To define the GNC in a neighborhood of \(B\), we need to extend the null surface \(N^+\) smoothly ``a little'' to the past of \(B\), and similarly extend \(N^-\) to the future of \(B\). We assume, henceforth, that this has been done.} Since we have two null surfaces we obtain two different GNCs, one based on \(N^+\) which we denote by \((u_+,v_+,x^A_+)\) and another based on \(N^-\) denoted by \((u_-,v_-,x^A_-)\). In general, these two coordinate systems will not agree in a neighborhood of \(B\) and will be related by a coordinate transformation that preserves neither GNC. We will not need the explicit form of the transformations between these coordinates but we note that (by construction)
\be\label{eq:coord-cont}
    \lb. (u_\pm, v_\pm) \rb\vert_B = (u=0,v=0)  \eqsp \lb. x^A_+\rb\vert_B = \lb. x^A_- \rb\vert_B
\ee
and
\be\label{eq:l-n-cont}
    \ell^a \equiv \partial_v = \lb. \partial_{v_+} \rb\vert_B = \lb. \partial_{v_-} \rb\vert_B \eqsp n^a \equiv \partial_u = \lb. \partial_{u_+} \rb\vert_B = \lb. \partial_{u_-} \rb\vert_B 
\ee
The spacetime metric \(g_{ab}\), which we assume is smooth written in either coordinate system, coincides at \(B\).

We define the \(3\)-volume elements \(\varepsilon^\pm_{abc}\) on \(N^\pm\) and the \(2\)-area elements \(\varepsilon^\pm_{ab}\) on the cross-sections of \(N^\pm\) as follows:
\be\label{eq:vol-conv}\begin{aligned}
    \text{on } N^+: \quad \varepsilon^+_{abc} &= n^d \varepsilon_{dabc} \eqsp&& \varepsilon^+_{ab} = - \ell^c \varepsilon^+_{cab} = - \ell^c n^d \varepsilon_{dcab} \\
    \text{on } N^-: \quad \varepsilon^-_{abc} &= - \ell^d \varepsilon_{dabc} \eqsp&& \varepsilon^-_{ab} = - n^c \varepsilon^-_{cab} = n^c \ell^d \varepsilon_{dcab}
\end{aligned}\ee
Note that on \(N^+\) these conventions are the same as those of \cite{CFP} while on \(N^-\) the sign of \(\varepsilon^-_{abc}\) is the opposite of that in \cite{CFP}. We have chosen these conventions so that the area elements on the bifurcation edge \(B\) induced from \(N^\pm\) coincide, that is,
\be\label{eq:area-B-conv}
    \lb. \varepsilon^+_{ab} \rb\vert_B = \lb. \varepsilon^-_{ab} \rb\vert_B
\ee\\

Similar to the case of a general null surface, we can now identify the boundaries of any two causal diamonds in any two spacetimes by identifying the GNCs \((u_\pm, v_\pm, x^A_\pm)\). Note that with this identification the bifurcation edge \(B \equiv (u=0,v=0)\) is common to all causal diamonds and is \emph{universal}, but the corners \(u = U(x^A)\) and \(v = V(x^A)\) depend on the specific choice of causal diamond and spacetime metric, and are thus not universal. Henceforth we will work with the covariant phase space $\bar{\mathscr{F}}$ of general relativity at the boundary $N$ of a causal diamond where the bifurcation edge \(B\) is a common universal surface for all spacetimes in \(\bar{\ms F}\).

\subsection{Reduced symmetry algebra \(\mf g_{\rm CD}\) at causal diamonds}
\label{sec:symm-CD}

Since the bifurcation edge \(B\) is universal the symmetry algebra for a causal diamond must preserve \(B\). Consider the null boundary symmetry algebra on the future null surface \(N^+\). From the form of the vector fields \(\xi^a\) in \cref{eq:symm-GNC} we see that the symmetries on \(N^+\) which preserve the surface \(B \equiv (v= 0)\) are the ones which satisfy \(\alpha(x^A)\vert_{N^+} = 0\). In other words, the bifurcation edge \(B\) breaks the affine supertranslation symmetry. Similarly, the affine supertranslations of the symmetry algebra at the past surface \(N^-\) are also broken.

A priori it seems we have two independent symmetries for the causal diamond: one induced from \(N^+\) and the other from \(N^-\), with the respective affine supertranslations set to vanish. However, there is a natural isomorphism between the future and past symmetries which follows from the smoothness of the vector field \(\xi^a\) in spacetime. To see this let \(\xi^a\) be a smooth vector field in the spacetime \(M\) which is a representative of a symmetry on \(N^\pm\) respectively, preserving the bifurcation edge \(B\). In the GNCs \((u_\pm, v_\pm, x^A_\pm)\) based on the null surfaces \(N^\pm\) we have (see \cref{eq:symm-GNC,eq:l-n-cont})
\be\begin{aligned}
    \xi^a &\equiv \beta_+ ( - v_+\partial_{v_+} + u_+ \partial_{u_+}) + X_+^A \partial_{A_+} + \ldots \\
    &\equiv \beta_- ( - u_- \partial_{u_-} + v_- \partial_{v_-}) + X_-^A \partial_{A_-} + \ldots
\end{aligned}\ee
where as before \(\ldots\) denotes the subleading terms in the respective GNCs. Note that while the GNCs do not match in a neighborhood of \(B\), from \cref{eq:coord-cont,eq:l-n-cont} and the smoothness of \(\xi^a\) at \(B\) we can conclude that
\be\label{eq:symm-match}
    \lb. \beta_+ \rb\vert_B = - \lb. \beta_- \rb\vert_B \eqsp \lb. X_+^A \rb\vert_B = \lb. X_-^A \rb\vert_B
\ee
This implies a natural isomorphism between the symmetry algebras on \(N^+\) and \(N^-\) given by the matching conditions \cref{eq:symm-match} at \(B\). Thus the elements of the symmetry algebra \(\mf g_{\rm CD}\) on the boundary of a causal diamond are given by \((\beta, X^A)\); for definiteness we choose \((\beta, X^A) = (\beta_+, X^A_+) = (-\beta_-, X^A_-)\) to represent an element in \(\mf g_{\rm CD}\).

The Lie brackets of the algebra \(\mf g_{\rm CD}\) can be derived from \cref{eq:bracket} by setting \(\alpha_1 = \alpha_2 = 0\). We have
\be\label{eq:CD-bracket}\begin{aligned}
    &\lb[ (\beta_1, X^A_1), (\beta_2, X^A_2) \rb] = (\beta, X^A) \\
    \text{where} \quad 
    \beta &= X^A_1 \partial_A \beta_2 - X^A_2 \partial_A \beta_1 \\
    X^A &= \lb[ X_1, X_2 \rb]^A = X^B_1 \partial_B X^A_2 - X^B_2 \partial_B X^A_1
\end{aligned}\ee
If \(X^A_1 = 0\) then \(X^A = 0\) hence symmetries of the form \((\beta, 0)\) form an infinite-dimensional abelian Lie ideal \(\mf b\) of \emph{boost supertranslations}. Thus,
\be
    \mf g_{\rm CD} \cong {\rm diff}(\bb S^2) \ltimes \mf b
\ee
Further, if \(\beta_1 = \text{constant}\) and \(X^A_1 = 0\) then \(\beta = X^A = 0\), that is, the symmetries of the form \((\beta = \text{constant}, 0)\) commute with any element of \(\mf g_{\rm CD}\) and thus form a \(1\)-dimensional Lie subalgebra \(\mf b_0\) of central elements which we call \emph{boosts}.\footnote{The terminology ``boosts'' for elements of \(\mf b_0\) is motivated by the fact that if one considers a bifurcate Rindler horizon in Minkowski spacetime, instead of a causal diamond, then Lorentz boosts which preserve the Rindler horizon are precisely elements in \(\mf b_0\).} Consider the quotient \(\mf g_{\rm CD}/\mf b_0 \cong {\rm diff} (\bb S^2) \ltimes (\mf b/\mf b_0)\). Then \(\mf g_{\rm CD}\) has the structure of a \emph{central extension} of \(\mf g_{\rm CD}/\mf b_0\) by the abelian Lie algebra \(\mf b_0\); the fact that this is a \emph{non-trivial} central extension is shown in \cref{sec:central-ext}.\footnote{Note that if one eliminates the non-constant boost supertranslations, for instance by imposing a weakly isolated horizon structure when \(N\) is stationary, then the central extension also becomes trivial (see for instance Sec.~IV.B \cite{Ash-Bah}).}

We show in \cref{sec:central-charge} that the charges associated to the central elements in \(\mf b_0\) can be interpreted as providing a ``first law'' for causal diamonds.

\subsection{Behavior of the fields and charges near the corners}
\label{sec:corner}

In this section we analyze the behavior of the relevant fields and charges on \(N^+\) near the corner \(p^+\); similar results hold also for \(N^-\) at \(p^-\). Essentially, for causal diamonds in smooth spacetimes the behavior of the fields of interest near \(p^+\) is the same as that of a light cone in Minkowski spacetime with some subleading corrections away from \(p^+\). We invoke the results of \cite{corner0,corner1} below. The main consequence of interest for our purposes is that in the limit to the corner \(p^+\) along \(N^+\) we have
\be
    \mc Q_\xi \to 0 \eqsp i_\xi \df\Theta(\delta g) \to 0
\ee
for all symmetries \(\xi \in \mf g_{\rm CD}\), and all metric perturbations \(\delta g_{ab}\) which are smooth at \(p^+\) in any spacetime. The limit \(\mc Q_\xi \to 0\) near the corner ensures that the total flux \(\mc F_\xi\) associated to the symmetries through all of \(N^+\) is finite and is, in fact, equal to the charge at \(B\). As we will show in \cref{sec:conservation}, along with smoothness of the spacetime at \(B\), this gives infinitely many conservation laws between the incoming and outgoing fluxes through any causal diamond. The implication of the limit \(i_\xi \df\Theta(\delta g) \to 0\) is as follows: a perturbation of the total flux on \(N^+\) can be written as (see \cref{eq:gen-cond0})
\be\label{eq:gen-cond0-N+}
    \delta \mc F_\xi = \int_{N^+}\df{\omega}(g;\delta g, \lie_{\xi}g) + \int_B i_{\xi}\df{\Theta}(\delta g) - \int_{p^+} i_{\xi}\df{\Theta}(\delta g)
\ee
where the integral over the corner \(p^+\) should be interpreted as a limit of integrals on cross-sections of \(N^+\) which suitably limit to \(p^+\) as described below. Since \(i_\xi \df\Theta(\delta g) \to 0\) in the limit to \(p^+\) and \(i_\xi \df\Theta(\delta g)\vert_B = 0\) (as \(\xi^a\) is tangent to \(B\) for any symmetry in \(\mf g_{\rm CD}\)), we have
\be\label{eq:gen-cond-N+}
    \delta \mc F_\xi = \int_{N^+}\df{\omega}(g;\delta g, \lie_{\xi}g)
\ee
for all symmetries \(\xi \in \mf g_{\rm CD}\) and all perturbations \(\delta g_{ab}\) which are smooth at \(p^+\) and \(B\). Thus the flux \(\mc F_\xi\), viewed as a function on the covariant phase space on \(N^+\), generates a Hamiltonian flow associated to the symmetry $\xi^a$ (see \cref{eq:gen-cond}).

We now describe the arguments leading to the above described result, referring to \cite{corner0,corner1} for the details. Since the value of the GNC coordinate \(v\) at the corner is direction-dependent, \(v\) not a regular coordinate at \(p^+\). Similarly, the cross-sections \(v = \text{constant}\) do not limit to \(p^+\). Thus, to analyze the behavior of the symmetries and charges we need a coordinate system on \(N^+\) which is more suited to the structure near \(p^+\). In a neighborhood of \(p^+\) such a coordinate system can be constructed as follows. Consider the tangent space \(Tp^+\) at \(p^+\), and let \(y^i = (y^0, y^1,y^2,y^3)\) be coordinates in \(Tp^+\) so that \(y^i\vert_p^+ = 0 \) and the coordinate vector fields \(\partial_i\) are orthonormal, with \(\partial_0\) being timelike and future-directed. Defining
\be
    r^2 = (y^1)^2 + (y^2)^2 + (y^3)^2 \eqsp u =  y^0 - r
\ee
the past-directed light cone in \(Tp^+\) from \(p^+\) is then given by \(u = 0\), and coordinatized by \((r,x^A)\) where \(x^A\) are coordinates on the space of past-directed null directions at \(p^+\) isomorphic to \(\bb S^2\).

There exists an exponential map from \(Tp^+\) to a sufficiently small neighbourhood of \(p^+\) so that \(y^i\) are coordinates in this neighborhood, called \emph{Riemann normal coordinates}. In such a neighborhood, using \((u,r,x^A)\) as coordinates, the metric \(g_{ab}\) takes the form (see \cite{corner0,corner1}, note we have changed some signs to conform to our orientation conventions)
\be\label{eq:g-corner}
    ds^2 = \mu du^2 - 2 \nu du dr - 2 \nu_A du dx^A + q_{AB} dx^A dx^B
\ee
The analysis of \cite{corner0,corner1} then shows that near the corner \(p^+\) the metric components in \cref{eq:g-corner} behave as
\be
    \mu = 1 + O(r^2) \eqsp \nu = 1 + O(r^4) \eqsp \nu_A = O(r^3) \eqsp q_{AB} = r^2 q_{AB}^0 + O(r^4)
\ee
where \(q_{AB}^0\) is the unit-metric on \(\bb S^2\). Here, for any tensor \(T_{AB\ldots}\) we use \(O(r^k)\) to denote that \(T_{AB\ldots} = r^k t_{AB \ldots}\) for some \(t_{AB\ldots}\) which, in general, has a non-vanishing limit as a tensor field on \(\bb S^2\) as \(r \to 0\). Roughly speaking, to leading order the metric \(g_{ab}\) near \(p^+\) behaves as the Minkowski metric at the corner of a light cone.

The expansion and shear of \(N^+\) have the behavior
\be
    \theta = -\frac{2}{r} + O(r^3) \eqsp \sigma_{AB} = O(r^3)
\ee
The normal to the foliation by \(r = \text{constant}\) surfaces is \(\hat n_a = dr\). The \Hajicek\ rotation \(1\)-form on \(N^+\) relative to the foliation by \(r\) is essentially the quantity denoted by \(\xi_A\) in \cite{corner0,corner1}, which satisfies
\be
    \hat \omega_A = O(r^2)
\ee
We have put a ``hat'' on the rotation \(1\)-form to emphasize its dependence on the foliation.\\

To consider the limit of the charges associated to the symmetries on \(N^+\), we now relate these coordinates to the GNC used in the main arguments above. The non-affinity \(\hat\kappa\) of the null generator \(\hat \ell^a \equiv - \partial_r\) is given by
\be
    \hat\kappa = - \partial_r \ln \nu = O(r^3)
\ee
and thus \(\hat \ell^a\) is an affine null generator of \(N^+\) up to \(O(r^3)\). Thus near \(p^+\), we can identify \(\hat \ell^a\) with the GNC null generator \(\ell^a \equiv \partial_v\) and the coordinate \(r\) with GNC coordinate \(v\) as
\be\label{eq:l-conv}
    \ell^a = \hat \ell^a + O(r^3) \eqsp v - V(x^A) = - r + O(r^4)
\ee 
Note that, as is to be expected, the cross-sections of \(N^+\) given by \(r = \text{constant}\) and those given by \(v = \text{constant}\) do not coincide. In particular, their normals \(\hat n_a \equiv dr\) and \(n_a \equiv - dv\) are related by
\be\label{eq:n-conv}
    n_a = \hat n_a + \partial_A V dx^A + O(r^3)
\ee

Now consider a symmetry \(\xi^a = -v \beta \partial_v + X^A \partial_A\) in GNC where, as before, \(X^A\) is tangent to the \(v = \text{constant}\) cross-sections. We rewrite this vector field as \(\xi^a = \hat f \hat \ell^a + \hat X^A \partial_A\) so that \(\hat X^A\) is tangent to the \(r = \text{constant}\) cross-sections. From \cref{eq:l-conv,eq:n-conv} we have
\be\label{eq:xi-corner}
    \hat f = -(V - r)\beta - X^A \partial_A V + O(r^3) \eqsp \hat X^A = X^A + O(r^3)
\ee
and also
\be\label{eq:betaX-corner}
    \beta_{\hat X} =  - \hat X^A \hat q_A{}^b \lie_{\hat \ell} \hat n_b = O(r^2)
\ee

The WZ charge \(\mc Q_\xi\) (see \cref{supcharge,diffcharge-alt}) evaluated on some cross-section \(S_r\) with \(r = \text{constant}\) is then
\be
    \mc Q_\xi[S_r] = \frac{1}{8\pi} \int_{S_r} \hat{\df\varepsilon}_2~ \lb[ \hat f \theta + \beta + \hat\beta_{\hat X} - \hat\omega_A \hat X^A \rb] 
\ee
From \cref{eq:xi-corner,eq:betaX-corner} we see that
\be
    \mc Q_\xi[p^+] = \lim_{r \to 0} \mc Q_\xi[S_r] = 0
\ee
where we have used that \(\hat{\df\varepsilon}_2 = r^2 \df\varepsilon_2^0 + O(r^4)\) with \(\df\varepsilon_2^0\) the area-element of the unit-sphere.

Next consider the integral of \(i_\xi \df\Theta(\delta g)\) on the cross-sections \(S_r\)
\be
    \int_{S_r} i_\xi \df \Theta(\delta g) = - \frac{1}{16\pi} \int_{S_r}\hat{\df\varepsilon}_2~ \hat f \left( \sigma^{AB} - \tfrac{1}{2}q^{AB}\theta \right) \hat q_A{}^a \hat q_B{}^b \delta g_{ab}
\ee
Any metric perturbation \(\delta g_{ab}\) which is smooth at \(p^+\) has smooth components in the Riemann normal coordinates \(y^i\) described above, and its spherical components behave as \(\hat q_A{}^a \hat q_B{}^b \delta g_{ab} = O(r^2)\). Thus, we have
\be
    \lim_{r \to 0} \int_{S_r} i_\xi \df\Theta(\delta g) = 0
\ee

\section{Conservation laws at causal diamonds}
\label{sec:conservation}

We now show that there exist an infinite-number of conservation laws associated to the symmetry algebra \(\mf g_{\rm CD}\) between fluxes through \(N^-\) and \(N^+\) for any causal diamond. These conservation laws follow directly from the smoothness of the relevant fields at the bifurcation edge \(B\).

First we show that the smoothness of the spacetime at \(B\) implies that the charges corresponding to the symmetries in \(\mf g_{\rm CD}\) evaluated at \(B\) are equal. From \cref{supcharge,diffcharge,eq:symm-match}, the charges at \(B\) induced from \(N^\pm\) are
\be\label{eq:charges-B}\begin{aligned}
    \mc Q_\xi[B] & = + \frac{1}{8\pi} \int_B \df\varepsilon_2^+~  \lb( + \beta - \omega_A^+ X^A \rb) \\
    \mc Q_\xi[B] & = - \frac{1}{8\pi} \int_B \df\varepsilon_2^-~  \lb( - \beta - \omega_A^- X^A \rb)
\end{aligned}\ee
where the difference in the sign of these expressions is due to our conventions for the area elements on \(N^\pm\) given in \cref{eq:vol-conv} and the matching conditions on the symmetries \cref{eq:symm-match}. To show that these charges are equal we need to consider the relation between the \Hajicek\ rotation \(1\)-forms \(\omega_A^+\) and \(\omega_A^-\) which can be obtained as follows. Let \(\ell^a_\pm\) and \(n^a_\pm\) be the extensions in the respective GNCs of the null vector fields \(\ell^a\) and \(n^a\) on \(B\). Then we can compute
\be\begin{aligned}
    \lb. \omega_A^+ \rb\vert_B & = - (q^+)_A{}^c n_b^+ \nabla_c^+ \ell^b_+ = (q^+)_A{}^c \ell_b^+ \nabla_c^+ n^b_+ \\
    & = (q^+)_A{}^c \ell_b^- \nabla_c^+ n^b_- \\
    & = (q^-)_A{}^c \ell_b^- \nabla_c^- n^b_- \\
    & = - \lb. \omega_A^- \rb\vert_B
\end{aligned}\ee
where in the first line we have used \(n_a^+ \ell^a_+ \vert_B = n_a \ell^a = -1\), in the second line we have used the fact that \(\ell^a\) and \(n^a\) are continuous at \(B\) (see \cref{eq:l-n-cont}), in the third line the continuity of the induced metric \(q_{ab}\) and the spacetime derivative operator \(\nabla\) (which follows from the smoothness of the metric \(g_{ab}\)) and in the last line the definition of \(\omega_A^-\) at \(B\). Thus, from the smoothness of the spacetime metric and the continuity of the GNCs at \(B\) we have\footnote{In the Newman-Penrose notation \cite{NP} \cref{eq:rotation-match} is simply the identity \(\beta + \bar\alpha = - (-\beta - \bar\alpha)\), while in the Geroch-Held-Penrose notation \cite{GHP} it is \(\beta - \bar\beta' = - (-\beta + \bar\beta')\), which follow from the continuity of the spin-coefficients of the spacetime derivative operator \(\nabla\) at \(B\).}
\be\label{eq:rotation-match}
    \lb. \omega_A^+ \rb\vert_B = - \lb. \omega_A^- \rb\vert_B
\ee
Combining \cref{eq:rotation-match} with \cref{eq:area-B-conv,eq:symm-match,eq:charges-B} we have
\be\label{eq:Q-match}
    \mc Q_\xi[B] \text{ from } N^+ = \mc Q_\xi[B] \text{ from } N^-
\ee

Next, we consider the fluxes through \(N^\pm\) given by
\be\begin{aligned}
    \mc F_\xi[N^+] &= \mc Q_\xi[B] - \mc Q_\xi[p^+] \\
    \mc F_\xi[N^-] &= \mc Q_\xi[B] - \mc Q_\xi[p^-]
\end{aligned}\ee
Note that the flux on \(N^+\) is \emph{outgoing} while that on \(N^-\) is \emph{incoming} relative to the causal diamond (in accordance with our conventions \cref{eq:vol-conv}). As shown in \cref{sec:corner} the charges at the corners \(p^\pm\) vanish and thus from \cref{eq:Q-match} we have
\be\label{eq:conservation}\begin{aligned}
    \mc F_\xi[N^+] & = \mc F_\xi[N^-] 
\end{aligned}\ee
That is, the \emph{incoming} flux through \(N^-\) is equal to the \emph{outgoing} flux through \(N^+\) for any symmetry in \(\mf g_{\rm CD}\). Thus, there are infinitely-many conservation laws associated to the symmetry algebra on any causal diamond in any spacetime in general relativity.\\

\begin{remark}[Affine supertranslations]\label{rem:affine}
    Note that in \cref{sec:symm-CD} we eliminated the affine supertranslations \(\alpha(x^A) \neq 0\) from the symmetry algebra of the causal diamond by demanding that the bifurcation surface \(B\) be preserved under the symmetries. If we had kept \(\alpha\) then \(i_\xi \df\Theta(\delta g)\vert_B \neq 0\) --- since such vector fields are not tangent to \(B\) --- and thus, the flux of the affine supertranslations is not a Hamiltonian generator on the phase space of the null boundary. Furthermore, the affine supertranslations \(\alpha_+ \ell^a\) defined on \(N^+\) and \(\alpha_- n^a\) defined on \(N^-\) cannot be matched at \(B\), as the corresponding vector fields are not continuous. Even if one imposes the condition \(\alpha_+(x^A) = \alpha_-(x^A)\) by hand, the charges corresponding to the affine supertranslations at \(B\) (see \cref{supcharge}) do not match since the expansions \(\theta_\pm\) along \(N^\pm\) need not be equal at \(B\) in general. Thus, there do not exist any conservation laws at a causal diamond in general spacetimes analogous to \cref{eq:conservation} for the affine supertranslations.
\end{remark}

\begin{remark}[Non-affine parametrization of the null generators]
    For convenience we chose the null generators of \(N^\pm\) to be affinely-parametrized, but our result is invariant under this choice. One can construct a GNC on a null surface relative to an arbitrarily parametrized null generator (with \(\kappa \neq 0\)). The resulting symmetry algebra is then as described in \cite{CFP} and \cref{sec:CFP-summ}. The affine supertranslations are eliminated by the condition \(f\vert_B = 0\), in which case the boost supertranslations in \(\mf b\) are parametrized by the function \(- (\lie_\ell + \kappa)f\) which is invariant under arbitrary rescalings of the null generators. For the boosts in \(\mf b_0\) we have \(- (\lie_\ell + \kappa)f = \text{constant}\). The remainder of our analysis can also be generalized in a similar fashion; we only note that since \(f\vert_B = 0\), the non-affinities \(\kappa_\pm\) of the generators of \(N^\pm\) do not enter into the matching of the symmetries and charges at \(B\) and the resulting conservation laws.
\end{remark}

\section{Central charges and area of the bifurcation edge}
\label{sec:central-charge}

As discussed in \cref{sec:symm-CD}, the symmetry algebra \(\mf g_{\rm CD}\) at the boundary of the causal diamond can be viewed as a non-trivial central extension of \(\mf g_{\rm CD}/\mf b_0\) by the \(1\)-dimensional abelian subalgebra \(\mf b_0\) of boosts. One expects the charges of such central elements to be of special significance. We show below that these charges are directly related to the area of \(B\), in analogy with the Wald entropy formula for black holes \cite{IW-noether-entropy}. A similar result was found in \cite{Donnelly:2016auv} through different considerations.

From the results of \cref{sec:conservation} we have that\footnote{It can be verified that for a causal diamond in Minkowski spacetime where \(B\) is a sphere of radius \(R\), our conventions \cref{eq:vol-conv} give \(\int_B \df\varepsilon^+_2 = {\rm Area}(B) = + 4\pi R^2\).}
\be\label{eq:boost-charge}
    \mc F_\beta[N^+] = \mc Q_\beta[B] = \frac{\beta}{8\pi} {\rm Area}(B)
\ee
for any \((\beta = \text{constant}, X^A = 0) \in \mf b_0\). This can be written in a more illuminating form as follows: consider the vector field \(\xi^a\vert_{N^+} = - v\beta \ell^a\) corresponding to an element in \(\mf b_0\), so that
\be
    \xi^b \nabla_b \xi^a = \kappa_{(\beta)} \xi^a \eqsp \kappa_{(\beta)} = - \beta = \text{constant} 
\ee
and thus
\be\label{eq:boost-charge1}
    \mc F_\beta[N^+] = \mc Q_\beta[B] = - \frac{\kappa_{(\beta)}}{2\pi} \times \frac{1}{4}{\rm Area}(B) 
\ee
If we interpret the charge \(\mc Q_\beta[B]\) as an ``energy'', the \(\tfrac{1}{4}{\rm Area}(B)\) as an ``entropy'' and \(- \frac{\kappa_{(\beta)}}{2\pi}\) as a ``temperature'', relative to the vector field \(\xi^a\), then \cref{eq:boost-charge1} takes the form of a ``first law'' \cite{Wald-book,IW-noether-entropy,Gourgoulhon:2005ng}. Note that if \(\xi^a\) is future-directed on \(N^+\) we have \(\beta < 0\) and so \(\kappa_{(\beta)} > 0\) and the temperature is negative. This difference in sign compared to the temperature of bifurcate Killing horizons essentially arises due to the fact that the future-directed null generator \(\ell^a\) points ``inwards'' on \(N^+\). Such a negative temperature was also found for causal diamonds in maximally symmetric spacetimes in \cite{Jacobson:2018ahi}. Also note that for asymptotically-flat stationary black holes the scaling of the horizon Killing vector field \(K^a\) is fixed by the requirement that at spatial infinity \(K^a\) asymptotes to a future-directed unit-normalized timelike Killing vector field (plus a rotational vector field). This completely fixes the scaling of the surface gravity, and hence the temperature, of the black hole. In contrast, there is no natural normalization for the boost vector fields at a causal diamond so we get an entire \(1\)-dimensional family of surface gravities \(\kappa_{(\beta)}\) and temperatures corresponding to the symmetries \(\mf b_0\).

The charges \(\mc Q_\beta[B]\) associated to elements of \(\mf b_0\) are central even in the sense of a Poisson bracket on the phase space of \(N^+\), i.e. the boost charges Poisson-commute with all the other charges. This can be seen as follows: since the fluxes \(\mc F_\xi\) are Hamiltonian on the phase space of \(N^+\) (see \cref{eq:gen-cond-N+}), we can define their Poisson bracket as (see also Sec.~8 \cite{CFP})
\be
    \{ \mc F_{\xi_1}, \mc F_{\xi_2} \} = - \int_{N^+} \df\omega(g; \lie_{\xi_1}g, \lie_{\xi_2}g)
\ee
for any two symmetries \(\xi_1,\xi_2 \in \mf g_{\rm CD}\). Since \(i_\xi \df\Theta\vert_B = 0\) and \(\mc F_\xi[N^+] = \mc Q_\xi [B]\) for any such symmetry it follows from the analysis of Sec.~8 \cite{CFP} that
\be\label{pb}
    \{ \mc Q_{\xi_1}[B], \mc Q_{\xi_2}[B] \} = \mc Q_{[\xi_1,\xi_2]}[B] - \int_B \lie_{\xi_1} \df{\mc Q}_{\xi_2} 
\ee 
where in the final term \(\df{\mc Q}_\xi\) is the \(2\)-form whose integral on \(B\) gives the charges \cref{eq:charges-B}. If this term is vanishing then the Poisson algebra of charges is isomorphic to the Lie algebra of spacetime symmetries in \(\mf g_{\rm CD}\) (see Sec.~8 \cite{CFP} for details). In the present case this term does indeed vanish; we have
\be
    \int_B \lie_{\xi_1} \df{\mc Q}_{\xi_2} = \int_B \lb[ i_{\xi_1} d \df{\mc Q}_{\xi_2} + d \lb( i_{\xi_1} \df{\mc Q}_{\xi_2} \rb) \rb] = 0
\ee
where the first term vanishes since any \(\xi_1 \in \mf g_{\rm CD}\) is tangent to \(B\), while the second term vanishes upon integration over \(B\). Thus the Poisson algebra of the charges on \(B\) is isomorphic to the Lie algebra \(\mf g_{\rm CD}\). The boost charges \cref{eq:boost-charge1} associated to the central elements in \(\mf b_0\) are then also central charges on the phase space on \(N^+\) in the sense of the Poisson algebra.\\

We emphasize that the appearance of central charges in the above analysis is quite different from that of previous approaches. In particular, we work in full nonlinear general relativity at any causal diamond in any spacetime satisfying the vacuum Einstein equations without any restriction to ``near horizon'' geometries of stationary black holes (as done in \cite{Carlip:1998wz,Carlip:2017xne,Guica:2008mu}). Since the Einstein equations are not conformally invariant there is no conformal symmetry or Virasoro algebra at the causal diamond in the general case we have considered. In fact, since we have allowed for non-vanishing shear, the induced \(2\)-metrics on the cross-sections of the causal diamond are not conformally related (see \cref{eq:shear-exp-GNC}). We always work with smooth vector fields as representatives of the symmetries (as opposed to the singular vector fields considered in \cite{Haco:2018ske}). Furthermore, as discussed above, the Poisson algebra of the charges in our case is isomorphic to the Lie algebra of symmetries with no additional central extension (in contrast with \cite{Speranza:2017gxd, Haco:2018ske, Carlip:1998wz}). The central extension we obtain already exists in the structure of the spacetime symmetry algebra \(\mf g_{\rm CD}\).

\section{Discussion}
\label{sec:disc}

We studied the covariant phase space formalism at the boundaries of causal diamonds in vacuum general relativity. In suitable Gaussian null coordinates, we showed that one can identify all causal diamonds and their bifurcation edges across all spacetimes, and that the symmetry algebra at the null boundaries of the casual diamond takes the form $\mathfrak{g}_{\text{CD}} \cong \text{diff}(\mathbb{S}^2)\ltimes \mathfrak{b}$ where \({\rm diff}(\bb S^2)\) maps different null generators of the causal diamond boundary into each other and $\mathfrak{b}$ consists of angle-dependent rescalings of the affine parameter along the null generators. Suitable smoothness conditions at the corners of the causal diamond imply that the Wald-Zoupas charges vanish at the corners --- so that the total flux across the null boundary is equal to the charge at the bifurcation edge --- and that the Wald-Zoupas fluxes define Hamiltonian generators of the symmetries on the null boundary phase space. The smoothness of the symmetry vector fields and the fields at the bifurcation edge then give rise to an infinite-number of conservation laws for the Wald-Zoupas fluxes between the past and future components of the causal diamond boundary. We also showed that the charge associated to the central elements of the symmetry algebra --- i.e. the elements of the subalgebra $\mathfrak{b}_0$ consisting of the angle-independent supertranslations --- is related to the area of the bifurcation edge through a ``first law'' similar to the Wald entropy formula for stationary black holes.\\

While our analysis focused on causal diamonds in classical vacuum general relativity we expect that it can be generalized to include matter fields described by a suitable QFT on curved spacetimes. For instance, it was shown in \cite{DPP} that a comparison of suitable states defined on causal diamonds in different spacetimes can be used to extract properties of the local curvature of the spacetimes. Similarly, in \cite{HI-news-info} it was shown that the relative entropy of quantum states in linearized general relativity in an asymptotically-flat black hole spacetime is related to the area of the black hole and Bondi flux at null infinity. The infinite-dimensional symmetry algebra at the causal diamond could be useful to analyze other properties of a QFT on curved spacetimes.

It has been conjectured that the conservation laws at null infinity strongly constrain the scattering matrix of quantum gravity in asymptotically-flat spacetimes \cite{Strominger:2013jfa}. Similarly, we expect the conservation laws derived in \cref{sec:conservation} can be used to constrain the transition amplitudes in quantum gravity on local causal diamonds. To do this one needs to suitably quantize the gravitational degrees of freedom on the null boundary (see for instance \cite{Wieland:2019hkz,Wieland:2017zkf, Wieland:2017cmf}) and promote the charges and fluxes to operators with the bracket structure \cref{pb} in the corresponding quantum theory. We leave further investigation of this problem to future work.

We also expect that our analysis can be extended to causal diamonds at an asymptotic boundary in a spacetime, an interesting example of which arises in the AdS/CFT duality. In this context, for an asymptotically-AdS spacetime, the entanglement entropy of a CFT state defined on a causal diamond lying on the asymptotic boundary (conformal to Minkowski spacetime with one fewer dimensions) is dual to the area of the Ryu-Takayanagi surface in the bulk spacetime \cite{Ryu:2006bv,Dong:2016eik}. The Ryu-Takayanagi surface can itself be considered as the bifurcation edge of an ``entanglement wedge''. Presumably, our analysis can be suitably generalized to this case, taking into account the asymptotic AdS boundary conditions. We expect that the resulting symmetries are related to the boundary modular Hamiltonian, and that the associated charges and fluxes could provide further insight into the bulk dual of boundary modular flow following \cite{Jafferis:2015del, Lashkari:2016idm, Bousso:2019dxk}.

\section*{Acknowledgements}
V.C. is supported in part by the Berkeley Center for Theoretical Physics; by the Department of Energy, Office of Science, Office of High Energy Physics under QuantISED Award de-sc0019380 and contract DE-AC02-05CH11231; and by the National Science Foundation under grant PHY1820912. K.P. is supported in part by NSF grants PHY-1404105 and PHY-1707800 to Cornell University.

\appendix

\section{Covariant phase space formalism and the Wald-Zoupas charges}
\label{sec:WZ}

The computation of boundary charges and their fluxes makes use of the covariant phase space formalism \cite{LW} and the Wald-Zoupas prescription \cite{WZ}. We quickly review the main ingredients needed for null boundaries, and refer the reader to \cite{CFP} for details.

Consider a diffeomorphism covariant theory of the metric \(g_{ab}\) as a dynamical field which is described by a Lagrangian $4$-form $\df{L}(g)$ that depends locally and covariantly on $g_{ab}$. Under perturbations $g \mapsto g + \delta g$ the Lagrangian changes as
\begin{align}
\delta \df{L} = \df E^{ab} \delta g_{ab} + d\df{\theta}(g; \delta g)
\end{align}
where \(\df E^{ab}\) is a \(4\)-form presenting the equations of motion of the theory and the $3$-form $\df{\theta}(g,\delta g)$ is the presymplectic potential. The \(3\)-form presymplectic current is defined by 
\begin{align}
\df{\omega}(g; \delta_1 g, \delta_2 g) = \delta_1 \df{\theta}(g; \delta_2 g) - \delta_2 \df{\theta}(g; \delta_1 g)
\end{align}
where $\delta_1 g$ and $\delta_2 g$ are any two independent perturbations.

Given a vector field $\xi^a$, one can then show that
\begin{align}\label{eq:omega-Q}
\df{\omega}(g; \delta g, \lie_{\xi}g) = d[\delta \df{Q}_{\xi} - i_{\xi}\df{\theta}(\delta g)]
\end{align}
where the \(2\)-form \(\df Q_\xi\) is the Noether charge \cite{LW, IW-noether-entropy, WZ}.

Consider now a null boundary $N$ in the spacetime and a spacelike hypersurface $\Sigma$ that intersects $N$ at some cross-section $S$. Integrating \cref{eq:omega-Q} we get
\begin{align}\label{eq:omega-Q-int}
    \int_{\Sigma}\df{\omega}(g; \delta g, \lie_{\xi}g) = \int_{S} \delta \df{Q}_{\xi} - i_{\xi}\df{\theta}(\delta g)
\end{align}
Two vector fields \(\xi^a\) and \(\tilde\xi^a\) are equivalent representatives of symmetries on \(N\) if \(\xi^a\vert_N = \tilde \xi^a\vert_N\) and the right-hand-side of \cref{eq:omega-Q-int} evaluated with \(\xi^a\) and \(\tilde\xi^a\) are equal for all backgrounds \(g \in \bar{\ms F}\), all perturbations \(\delta g\) within \(\bar{\ms F}\) and all cross-sections \(S\) of \(N\). The boundary symmetries on \(N\) are then given by vector fields \(\xi^a\) factored out by the above equivalence relation.

From the above identity it would be ``natural'' to define a charge at \(S\) associated to a symmetry \(\xi^a\) as a function \(Q_\xi[S]\) on phase space so that 
\be\label{eq:charge}
    \delta Q_\xi[S] = \int_{S} \delta \df{Q}_{\xi} - i_{\xi}\df{\theta}(\delta g)
\ee
for all perturbations \(\delta g\) within \(\bar{\ms F}\) and all cross-sections \(S\). Unfortunately, in general, the right-hand-side of \cref{eq:charge} is not integrable in phase space and no function \(Q_\xi\) satisfying \cref{eq:charge} exists on the phase space. As shown in \cite{WZ} the integrability condition for the existence of a charge \(Q_\xi[S]\) for some symmetry \(\xi^a\) is
\begin{align}
    0 = (\delta_1 \delta_2 - \delta_2 \delta_1) Q_{\xi} = -\int_{S}i_{\xi}\df{\omega}(g,\delta_1 g, \delta_2 g)
\end{align}
for all perturbations \(\delta_1 g, \delta_2 g\) within \(\bar{\ms F}\) and all cross-sections \(S\). This above criteria is not satisfied, even at null infinity in general relativity, except in very special cases \cite{WZ}.

Nevertheless, Wald and Zoupas \cite{WZ} developed a prescription for defining a modified charge which is always integrable. Define a boundary presymplectic potential $\df{\Theta}(g;\delta g)$ for the pullback to $N$ of the presymplectic current, 
\begin{align}\label{eq:Theta-defn}
\bar{\df{\omega}}(g,\delta_1 g, \delta_2 g) = \delta_1 \df{\Theta}(g,\delta_2 g) - \delta_2 \df{\Theta}(g,\delta_1 g)
\end{align}
where $\bar{\df{\omega}}$ denotes the pullback to \(N\). Then define the \emph{Wald-Zoupas charge} (WZ charge) by 
\begin{align}\label{eq:WZ charge} 
\delta \mc Q_{\xi}[S] = \int_{S} \delta \df{Q}_{\xi} - i_{\xi}\df{\theta}(\delta g) + i_{\xi}\df{\Theta}(\delta g)
\end{align}
It can be shown using \cref{eq:omega-Q-int,eq:Theta-defn} that \(\delta\mc Q_\xi[S]\) is integrable in phase space. Thus \cref{eq:WZ charge} determines a function $\mc Q_{\xi}[S]$ up to a constant of integration on $\bar{\mathscr{F}}$, which can be fixed by choosing a reference solution $g_0$ such that \(\mc Q_{\xi}[S]\big\lvert_{g_0}= 0\) for all symmetries $\xi^a$ and all cross-sections $S$.

The prescription given by Wald and Zoupas is to choose the \(3\)-form \(\df\Theta\) such that \(\df\Theta(g;\delta g)\) vanishes for all perturbation \(\delta g_{ab}\) for any background \(g_{ab}\) which is stationary, and to choose the reference solution \(g_0\) to also be stationary. The consistency conditions for such choices and the ambiguities in them are detailed in \cite{WZ}.

If the above choices can be made then the flux \(\mc F_\xi[\Delta N]\) of the WZ charge \(\mc Q_\xi\) through a part of the null boundary \(N\) is given by \cite{WZ}
\be
    \mc F_\xi[\Delta N] = \int_{\Delta N} \df{\Theta}(g; \lie_{\xi}g)
\ee
It can also be shown that the perturbed flux \(\delta\mc F_\xi\) for any symmetry \(\xi^a\) and any perturbation \(\delta g_{ab}\) satisfies (see Eq.~29 \cite{WZ})
\begin{align}\label{eq:gen-cond0}
    \delta \mc F_\xi = \int_{N}\df{\omega}(g;\delta g, \lie_{\xi}g) + \int_{\partial N}i_{\xi}\df{\Theta}(\delta g)
\end{align}
If $i_{\xi}\df{\Theta}(\delta g) \to 0$ on $\partial N$ for \emph{all} perturbations \(\delta g_{ab}\) then $\mc F_{\xi}$ is a function on the covariant phase space \(\bar{\ms F}\) satisfying
\be\label{eq:gen-cond}
    \delta \mc F_\xi = \int_{N}\df{\omega}(g;\delta g, \lie_{\xi}g)
\ee
for all perturbations \(\delta g_{ab}\), that is, \(\mc F_\xi\) defines a Hamiltonian which generates the flow on the covariant phase space $\bar{\mathscr{F}}$ associated to the symmetry $\xi^a$.\\

In \cite{WZ} this procedure was applied to the asymptotic symmetries at null infinity in general relativity to derive the charges and symmetries for the BMS algebra. The case of finite null boundaries in vacuum general relativity was handled in \cite{CFP}, where it as shown that the notion of symmetries defined below \cref{eq:omega-Q-int} coincides with those defined in \cref{sec:review} and the Wald-Zoupas prescription gives the charges and fluxes described in \cref{supcharge,diffcharge}. In \cite{CFP} the reference solution (used in the Wald-Zoupas prescription) was chosen to be the horizon of a Schwarzschild black hole in the limit that the mass tends to zero. It was shown that this reference solution satisfies all the criteria given by Wald and Zoupas \cite{WZ}. In this paper, we simply adopt the formulae for the charges and fluxes from \cite{CFP} and do not analyze the choice of reference solution in detail.

\section{Structure of \(\mf g_{\rm CD}\) as a central extension}
\label{sec:central-ext}

In this section we explore the structure of the summetry algebra \(\mf g_{\rm CD}\) of the causal diamond as a non-trivial extension of \(\mf g_{\rm CD}/\mf b_0\) by the boosts \(\mf b_0\).

Recall from \cref{sec:symm-CD} that the elements of \(\mf g_{\rm CD} \cong {\rm diff} (\bb S^2) \ltimes \mf b\) are of the form \((\beta , X^A)\) where \(\beta\) is a smooth function and \(X^A\) a vector feld on \(\bb S^2\). The central elements (i.e. those which commute with all other elements in \(\mf g_{\rm CD}\)) of boosts in \(\mf b_0\) are the ones given by \((\beta = \text{constant}, 0)\). Consider the quotient \(\mf g_{\rm CD}/\mf b_0\) which consists of equivalence classes given by the relation \((\beta, X^A) \sim (\beta + \text{constant}, X^A)\). Thus, \(\mf g_{\rm CD}\) is a central extension of \(\mf g_{\rm CD}/\mf b_0\) by the abelian algebra \(\mf b_0\). We now show that this central extension is, in fact, a non-trivial central extension. What this means is the following:

{\it Does the bracket of two representative elements belonging to \(\mf g_{\rm CD}/\mf b_0\), computed in \(\mf g_{\rm CD}\), have a non-vanishing \(\mf b_0\)-part?}

\noindent If the answer is `no' then the central extension is trivial and \(\mf g_{\rm CD}\) will be a direct product of \(\mf g_{\rm CD}/\mf b_0\) and \(\mf b_0\). If the answer is `yes' then \(\mf g_{\rm CD}\) has the structure of a non-trivial central extension of \(\mf g_{\rm CD}/\mf b_0\) by \(\mf b_0\).\footnote{In a more mathematical language, every central extension of \(\mf g_{\rm CD}/\mf b_0\) by the abelian algebra \(\mf b_0\) corresponds to a \(2\)-cocycle in the cohomology group \(H^2(\mf g_{\rm CD}/\mf b_0, \mf b_0)\) (see  Sec.~IV.2 \cite{Knapp}). Our computation in this section amounts to showing that the cocycle which gives the Lie algebra structure of \(\mf g_{\rm CD}\) is non-trivial.}

Since the symmetry algebra \(\mf g_{\rm CD}\) is independent of the metric \(q_{AB}\) on \(N\) the null boundary of the causal diamond, we can deduce its structure in any choice of metric, in particular it is convenient to choose \(q_{AB}\) to be the metric of a unit-sphere in the standard \((\theta,\phi)\) coordinates on \(\bb S^2\). We now compute the bracket of any two elements in \(\mf g_{\rm CD}/\mf b_0\), with the only relevant case being the bracket between an element of \(\mf b/\mf b_0\) with an element of \({\rm diff}(\bb S^2)\) which gives (see \cref{eq:CD-bracket})
\be
    [(\beta_1,0), (0, X^A_2)] = (\beta, 0) \quad\text{with}\quad \beta = - X^A_2 \partial_A \beta_1
\ee

To answer the above question we expand in terms of spherical harmonics \(Y_{l,m}(\theta,\phi)\). Note that elements of \(\mf b_0\) are purely \(l=0\) spherical harmonics. Now let \(\beta_1\) be a \(l_1\)-harmonic with \(l_1 \geq 1\) and \(X^A_2\) be a \(l_2\)-vector harmonic. We can write \(X^A_2 = \partial^A X + \epsilon^{AB} \partial_B \tilde X\) where \(X, \tilde X\) are \(l_2\)-harmonics with \(l_2 \geq 1\). We want determine whether \(\beta\) can have a non-trivial \(l = 0 \) part, that is a non-vanishing constant piece.

Before considering the general case, we note the following example
\be\label{eq:example}
    \beta_1 = \cos \theta \eqsp X^A_2 \equiv - \sin\theta \partial_\theta \implies \beta = \sin^2 \theta
\ee
Thus, \(\beta\) has non-vanishing \(l=0,2\) parts in terms of spherical harmonics. This already shows that \(\mf g_{\rm CD}\) is a non-trivial central extension of \(\mf g_{\rm CD}/\mf b_0\) by \(\mf b_0\).

For the general situation, first consider the case \(X = 0\), \(X^A_2 = \epsilon^{AB} \partial_B \tilde X\). Then we have, by integrating-by-parts on the unit-sphere (we leave the area element implicit for notational convenience)
\be\begin{aligned}
    \int \beta~ \bar Y_{l=0,m} \propto \int \beta = - \int \epsilon^{AB} \partial_B \tilde X \partial_A \beta_1 = - \int \partial_B (\epsilon^{AB} \tilde X \partial_A \beta_1) = 0
\end{aligned}\ee
Thus, \(\beta\) always has a vanishing \(l=0\) component for \(X^A_2 = \epsilon^{AB}\partial_B \tilde X\).

Next consider the case \(\tilde X = 0\), \(X^A = \partial^A X\) we have
\be\begin{aligned}
    \int \beta~ \bar Y_{l=0,m} \propto \int \beta & = - \int \partial^A X \partial_A \beta_1 = \int X \partial^2 \beta_1 \\
    &  = - l_1(l_1 + 1) \int X \beta_1 
\end{aligned}\ee
Expanding the functions \(X\) and \(\beta_1\) in terms of the corresponding spherical harmonics, and using the orthonormality and completeness of the spherical harmonic basis we conclude that, the right-hand-side is non-vanishing if and only if 
\be\label{eq:L-constraints}\begin{aligned}
    l_1 = l_2 \geq 1 \eqsp m_2 = - m_1
\end{aligned}\ee
Thus, \(\beta\) has a non-vanishing constant (\(l = 0\)) part whenever \cref{eq:L-constraints} is satisfied, which has many solutions; taking \(l_1 = l_2 = 1\) and \(m_1 = m_2 = 0\) gives the above example \cref{eq:example}. Thus, \(\mf g_{\rm CD}\) is a non-trivial central extension of \(\mf g_{\rm CD}/\mf b_0\) by \(\mf b_0\) as we wished to show.



\bibliographystyle{JHEP}
\bibliography{FluxConservation}

\end{document}